\documentclass[twocolumn,showpacs,amsmath,amssymb,nofootinbib,pra,superscriptaddress]{revtex4-1}

\usepackage{comment}
\usepackage{graphicx} 
\usepackage{xcolor}

\usepackage[
colorlinks=true,
    linkcolor={red!50!black},
    citecolor={blue!50!black},
    urlcolor={blue!80!black}]{hyperref}

\begin{document}

\title{Real-Time Dynamics of an Impurity in an Ideal Bose Gas in a Trap}

\author{A.~G. \surname{Volosniev}}
\affiliation{Department of Physics and Astronomy, Aarhus University, DK-8000 Aarhus C,  Denmark}
\affiliation{Institut f{\"u}r Kernphysik, Technische Universit{\"a}t Darmstadt, 64289 Darmstadt, Germany}
\author{H.-W. \surname{Hammer}}
\affiliation{Institut f{\"u}r Kernphysik, Technische Universit{\"a}t Darmstadt, 64289 Darmstadt, Germany}
\affiliation{
ExtreMe Matter Institute EMMI, GSI Helmholtzzentrum f{\"u}r Schwerionenforschung GmbH, 64291 Darmstadt, Germany}
\author{N.~T. \surname{Zinner}}
\affiliation{Department of Physics and Astronomy, Aarhus University, DK-8000 Aarhus C,  Denmark}

\begin{abstract}
We investigate the behavior of a harmonically trapped system consisting of an impurity in a dilute ideal Bose gas after 
the boson-impurity interaction is suddenly switched on. As theoretical framework, we use a field theory approach in 
the space-time domain within the T-matrix approximation. We establish the form of the corresponding T-matrix and address 
the dynamical properties of the system. As a numerical application, we consider a simple system of a weakly interacting impurity 
in one dimension where the interaction leads to oscillations of the impurity density. 
Moreover, we show that the amplitude of the oscillations can be driven
by periodically switching the interaction on and off.
\end{abstract}


\maketitle

\section{Introduction}

Understanding the behavior of an impurity in a host medium is a fundamental question in 
quantum many-body physics. 
On the one hand, this is motivated by the possibility to use an 
impurity particle as a probe in experiments. On the other hand,
such systems help to gain insights and build analytical and numerical tools 
to approach more complicated impure systems theoretically.   
A standard example is an electron placed in an ionic crystal, the solution to which 
can be approached by introducing the concept of the polaron - a quasiparticle that can be 
visualized as an electron dressed by collective excitations in the medium. This concept 
was first proposed in 1933 by Lev Landau \cite{landau1933} and has been widely used ever 
since \cite{fronlich1954,feynmanbook,devreese2009}. 
 One particularly interesting question in the study of impure systems is the time dynamics of an impurity in a medium. This question was mainly addressed for charged impurities in polar lattices or organic crystals \cite{mott1977, economou1992, Silinsh1995, kenrow1996, Ge09011998,swanson2000, ku2007, fehske2011, trugman2012} and still represents a challenge for the theoretical description.

Modern experiments with cold atomic gases produce systems where the density of immersed 
atoms is much smaller than the density of the host medium which can consist of fermionic 
\cite{zwierlein2009,grimm2012} or bosonic particles
\cite{widera2012,oberthaler2013}. The scattering length between host and impurity particles 
can be tuned using Feshbach resonances and the spatial geometry of the system can be 
designed using externally modulated traps  \cite{chin2010,kohl2009,kohl2012, catani2012, wenz2013}.
It is also worth to note hybrid set-ups which contain a single ion in a neutral system 
\cite{schmid2010,kohl2010} or an electron  in a Bose-Einstein condensate (BEC) \cite{pfau2013},
which leads to a different shape of the interaction potential in the system. 

The experimental progress in the realization of impure cold atomic systems allows to investigate 
the associated quasiparticles that appear due to the interaction between an impurity and the medium. 
In the literature these quasiparticles in a Fermionic (Bosonic) medium are often referred to as Fermi 
(Bose) polarons in analogy with Landau's concept. Experimental realization of degenerate imbalanced Fermi gases and subsequent observation of a polaron in a Fermi sea \cite{zwierlein2009}, drove a wave of new theoretical
activity in the Fermi polaron problem where a number of reliable results has been produced using different Monte Carlo 
algorithms \cite{lobo2006,prokofev2008,Bour:2014bxa}, variational wave functions \cite{chevy2006, zwerger2009} and T-matrix approaches \cite{chevy2007}. 
For a review of these activities, see Ref.~\cite{bruun2014} 
and references therein.

The experimental and theoretical study of an neutral impurity in a system of cold bosons has not yet 
been pursued as extensively as in the fermionic case. The theoretical 
approaches were motivated by systems 
of superfluid $^4$He doped with $^3$He, neutrons, etc. and 
were formulated for homogeneous systems~\cite{girardeu1961, pines1962}.
They have usually been applied to weakly-interacting dilute systems.  
Low-density cold 
atomic gases offer a new playground for theoretical studies. In these systems, the impurity-boson 
interaction can often be regarded of zero-range with a tunable coupling constant. The temperature of the medium 
can be very low and as a first approximation it can be put to zero. 

If the interaction in the system is weak, the usual approach to calculate observables
relies on mean-field equations, e.g. the Gross-Pitaevski equation. These equations have allowed the study of the 
motion of an impurity through a BEC in different spatial dimensions \cite{astrakharchik2004} and self-localization 
of an impurity in a trapped BEC \cite{blume2006}. Although the mean-field equations are limited in interaction strength 
they usually offer a natural way to include the effects of the trap which is present in experiments. To study regimes 
where the boson-impurity interaction is strong, one can use variational wave functions 
to extract an upper bound for the energy as well as the structure of the system 
\cite{timmermans2006,timmermans2006a,demler2014,dassarma2014, amin2014}.

In cases where the Bogoliubov approximation for the condensate mode is applicable, a commonly used approach 
is to map the original Hamiltonian onto the Fr{\"o}hlich Hamiltonian \cite{fronlich1954}.
For cold atomic gases this Hamiltonian was approached using the Feynman path integral formulation \cite{devreese2009a}, 
diagrammatic Monte Carlo \cite{devresee2014}, the renormalization group approach \cite{demler2014a}, and 
using a variational wave functions in the form of correlated Gaussians \cite{demler2014b}.  It was also shown 
that using the T-matrix approximation, the problem can be approached almost analytically to some extent~\cite{schmidt2013}. 
For time-independent homogeneous systems, these methods are able to predict the 
effective mass of the polaron for weak and strong coupling regimes.

The present work continues the study of an neutral impurity immersed in a Bose gas at zero temperature. 
Using the field-theoretical 
approach and the T-matrix approximation, we investigate the real-time dynamics of an impurity in a trapped 
ideal Bose gas after the impurity-boson interaction is suddenly turned on.  
We first work out the form of the in-medium T-matrix. This knowledge allows us 
to investigate the dynamical properties of the system. As a numerical application, 
we consider a weakly-interacting one-dimensional system. After the interaction is turned on, the impurity oscillates 
in the trap. We also show that by turning the interaction on and off periodically one can drive the amplitude of the oscillations.

The paper is organized as follows: in Section II we review the basics \cite{feynmanbook,abrikosovbook} of the field theory approach 
to many-body problems in the space-time domain with a particular emphasis on harmonically trapped systems and T-matrix approximation. 
In Sections III and IV, we derive the T-matrix for systems of
different spatial dimension analytically and discuss our approach to dynamical properties. This is illustrated by a numerical application to the non-trivial 
example of a weakly-interacting one-dimensional system. Our conclusions and a discussion of possible applications and extensions of this work
are given in Section V. Finally, we include four appendices with technical details and explicit derivations used in the main text.

\section{Formulation of the Problem}
{\it Hamiltonian}.
We consider a system with one impurity particle of mass $m_I$ and $N$ non-interacting bosons of mass $m_B$
in $D$ spatial dimensions. The Hamiltonian for this system consists of three pieces
\begin{align}
H=H_{B}+ H_{I}  + W,
\end{align}
where $H_{B}$ describes the system of bosons, $H_{I}$ is the 
Hamiltonian for the impurity, and $W$ is the impurity-boson interaction.
The ideal, harmonically trapped Bose gas is described by
the Hamiltonian
\begin{align}
H_B = \int \phi^{\dagger}(x) \left[-\frac{\hbar^2}{2m_B}
\frac{\partial^2}{\partial x^2}+\frac{m_B\Omega^2x^2}{2}\right]\phi(x) \mathrm{d}x,
\end{align}
whereas the Hamiltonian for the impurity has the form
\begin{align}
H_I =\int \psi^{\dagger}(x) \left[-\frac{\hbar^2}{2m_I}
\frac{\partial^2}{\partial x^2}+\frac{m_I\Omega^2x^2}{2}\right]\psi(x)  \mathrm{d}x,
\end{align}
where $x$ is a coordinate (scalar in one spatial dimension or vector in two or three spatial dimensions) and
$\phi (\psi)$ are the field operators that annihilate a boson (impurity). 
These operators are defined in a standard way, for instance $\phi(x)=\sum_{i\geq 0} b_i 
\Phi_i(x)$, where $b_i$ is the annihilation operator in second quantization and $\Phi_i(x)$ is 
the normalized one-body eigenfunction of $H_B$. For clarity, the field operators will 
be denoted with lower-case letters while the wave functions will be written with corresponding capital letters. 

The interaction between the impurity and the bosons is switched on at time $t=0$. It can be written
as
\begin{equation}
W=\theta(t)\int \phi^\dagger(x_1)
\psi^\dagger(x_2)V(x_1-x_2)\phi(x_1)\psi(x_2)\mathrm{d}x_1 \mathrm{d}x_2\; ,
\label{eq-int}
\end{equation} 
where $V$ is the impurity-boson interaction potential and $\theta(t)$ is the Heaviside step function.\footnote{
Note that the effective single-channel description implied by the interaction in 
Eq.~(\ref{eq-int}) can be used for broad Feshbach resonances \cite{chin2010}.}
Initially the system is in the non-interacting ground state, but the dynamics becomes 
non-trivial after the interaction is turned on at $t=0$.  Note, that introducing a trap makes the calculations much more inovolved compared to the homogeneous case. However, some aspects of the problem become more straightforward. For example to describe the homogeneous case one often works in a finite space parametrized by some length, $L$, and then takes the limit $N (L)\to\infty, N/L^D \to \mathrm{const}$. In the trapped case the system is localised and taking this limit is not required.  

{\it Wave function for impurity}. To describe dynamics of the impurity after for $t>0$, we focus on the following object
\begin{equation}
I(y,t)=\int F(y;z_1,...,z_N;t)\prod_{i=1}^N \Phi^{*}_0(z_i) \mathrm{d}z_i,
\end{equation}
where $F(y;z_1,...,z_N;t)$ is the wave function of the system of one impurity with coordinate $y$ and $N$ bosons with 
coordinates $z$ at time $t$. In the present paper, we consider a very dilute Bose gas for which the ladder approximation discussed 
below holds. In this approximation  the density of the bosons is not changed drastically, so the object $I(y,t)$ can be considered 
as the wave function of the impurity at time $t$, even though the impurity alone cannot be treated as an isolated system. 
If we know the wave function for the impurity at $t=0$, then $I(y,t)$ for $t>0$ can be obtained as 
\begin{equation}
I(y,t)=i\int\mathrm{d}x G_I(x,0,y,t)I(x,0),
\label{eq-wave-funct}
\end{equation}
where $G_I$ is the Green's function for the impurity
\begin{align}
G_I(x,t,x',t') = 
-i\frac{\langle  T(\psi_i(x',t') \psi_i^\dagger(x,t) S(\infty))\rangle }{\langle S(\infty)\rangle},
\label{eq:greenfirst}
\end{align}
defined using field operators in the interaction picture, i.e. $\psi_i(x,t)=e^{iH_I t/\hbar}\psi(x) e^{-iH_It/\hbar}$.
Here $T$ is the standard time ordering operator (larger times are always to the left); and 
$S(t)=T\exp\left(-\frac{i}{\hbar}\int_{-\infty}^t W_i(t')\mathrm{d}t'\right)$. The angle brackets in Eq.~(\ref{eq:greenfirst}) 
indicate averaging over the ideal Bose gas. As we consider only one impurity in a non-interacting Bose gas, 
${\langle S(\infty)\rangle}\equiv 1$. In the diagram technique this is very easily seen, since with 
one impurity and no boson-boson interactions only connected diagrams are possible. It is obvious that $G_I(x,t,x',t')$ should depend only on $t'-t$, since the presented formulation can also correspond to a system where the boson-impurity interaction is constant in time and the impurity is immersed in the medium at time $t$.

{\it Bogoliubov approximation}.
The non-interacting Bose gas at zero temperature has all particles in the ground state of the one-body Hamiltonian.  
To deal with macroscopic number of particles occupying one level, we 
utilize the so-called Bogoliubov prescription. First we write the field operators for
 the particles in the condensate and particles depleted from the condensate explicitly,
 i.e. $\phi(x,t)=b_0\Phi_0(x)+\phi'_i(x,t)$, where $\phi'_i(x,t)=e^{iH_Bt/\hbar}
\left[\sum_{k\neq0} b_k \Phi_k(x)\right]e^{-iH_Bt/\hbar}$. To write this 
expression down we assume that the energy of the condensate is zero, 
which for a harmonic trap technically corresponds to subtracting a constant term $D\hbar \Omega/2$ from $H_B$ and $H_I$. 
After the operators for the condensate and the depleted bosons are separated, 
we obtain  four interaction terms in Eq.~(\ref{eq-int}). In these four terms, assuming that the number of depleted bosons is small, 
we perform the substitution $b_0 \simeq b_0^{\dagger}\simeq \sqrt{N_0}$, 
where $N_0$ is the number of particles 
in the condensate. This number should be treated as a variable that minimizes the total energy.
We want to investigate the impurity in a dilute gas at zero temperature
assuming that the density of the condensate is not appreciably changed due to the impurity. This is done in the so-called ladder approximation 
with only one particle depleted from the condensate at any moment of time. 
So if the the total number of bosons is large, i.e. $N \gg 1$, we may substitute the operators $b_0$ and $b_0^\dagger$ with~$\sqrt{N}$.  

\begin{figure}
\centerline{\includegraphics[scale=0.5]{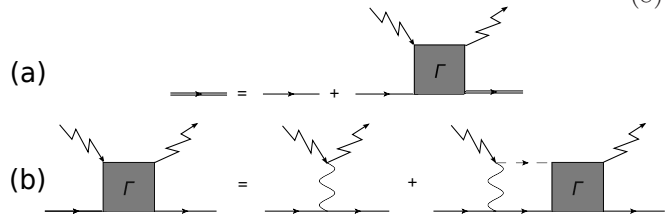}}
\caption{Diagrams representing main equations of the ladder approximation discussed in the text. (a) Eq.~(\ref{green}) 
for the impurity Green's function. (b) Eq.~(\ref{gamma}) for the in-medium T-matrix. Solid thick (thin) lines stand 
for the impurity Green's function, $G_I$ ($G_I^0$). The filled boxes represent the in-medium $T$-matrix, $\Gamma$. 
The wavy lines denote the interaction potential $V$. The zig-zag lines depicts the condensate particles, and the dashed 
line is the free Green's function for a boson depleted from the condensate, $G_B^0$.}
\label{fig:diagr}
\end{figure}

{\it Ladder approximation}.
We consider a dilute gas, such that $(n^{1/D}R)\ll 1$, where $R$ defines the  relevant interaction range.
The density $n$ is defined as the peak density of the BEC\footnote{Note that, for simplicity, we generically use the term BEC for a system with microscopic occupation number of the ground state.  This system, however, has different properties in different spatial dimensions \cite{mullin1997}. } in the trap. 
If  $(n^{1/D}R)\ll 1$, the system can be approached using the ladder approximation where only one boson is depleted at any moment of time
\begin{align}
G_I(x,t,x',t')=G^0_I(x,t,x',t')+  N\int \prod_{i=1,2}\mathrm{d}x_i \mathrm{d}y_i\mathrm{d}t_i \Phi_0(y_1)  \nonumber \\ 
\Phi_0^*(y_2)  G^0_I(x,t,x_1,t_1)  \Gamma(x_1,y_1,t_1,x_2,y_2,t_2) G_I(x_2,t_2,x',t'),
\label{green}
\end{align}
where $\Gamma$ is the in-medium T-matrix discussed below and $G^0_I$ is the free impurity propagator determined 
from Eq.~(\ref{eq:greenfirst}) with $W=0$. Note that in this paper the integration limits $\pm \infty$ are implied 
if nothing else is specified. It is worth to note that the integration over time variables is effectively always done over 
a finite interval, since both $G^0_I (x_1,t_1,x_2,t_2)$ and $\Gamma(x_1,y_1,t_1,x_2,y_2,t_2)$ vanish for $t_2<t_1$. 
This observation follows since the impurity cannot propagate backwards in time. The diagrammatic representation of 
Eq.~(\ref{green}) is presented in Fig.~\ref{fig:diagr}.  
The in-medium T-matrix, $\Gamma$, is written as   
an integral equation,
\begin{widetext}
\begin{align}
\Gamma(x_1,y_1,t_1,x_2,y_2,t_2) = &\delta (x_1-x_2)\delta(y_1-y_2)\delta(t_1-t_2) V(x_1-y_1)  
\nonumber \\ +&\frac{i}{\hbar} V(x_1-y_1) \int G^0_I(x_1,t_1,y,\tau)G^0_B(y_1,t_1,y',\tau)  
\Gamma(y,y',\tau,x_2,y_2,t_2) \mathrm{d}y\mathrm{d}y'\mathrm{d}\tau\; ,
\label{gamma}
\end{align}
\end{widetext}
where $G^0_B$ is the free Green's function for depleted bosons
(cf.~Fig.~\ref{fig:diagr}).
For the homogeneous case, the ladder approximation in the momentum-frequency domain 
was shown to capture the qualitative behavior of the impurity in the Bose-polaron problem~\cite{schmidt2013}.

{\it Free Green's functions in a harmonic trap}.
In a 1D harmonic trap, the free Green's function for $t>0$ 
can be written as 
\begin{equation}
G^0_I(x,0,x',t) = -i\sum_{n=0}^{\infty} \Psi^{*}_n(x)
\Psi_n(x')e^{-i{\epsilon_nt}/\hbar}\; ,
\label{eq:greenharm}
\end{equation}
with $\epsilon_n = \hbar\Omega n$ (recall that we subtract the zero point motion). 
The corresponding one-body wave function can be always chosen as a real function, i.e.
\begin{equation} \Psi_n(x) = 
\frac{e^{
- \frac{m_I\Omega x^2}{2 \hbar}}}{\sqrt{2^n\,n!}} \left(\frac{m_I\Omega}{\pi \hbar}\right)^{1/4} H_n\left(\sqrt{\frac{m_I\Omega}{\hbar}} x \right),
\label{eq:hermite}
\end{equation} 
where $H_n$ is a Hermite Polynomial of order $n$.  
From Eq.~(\ref{eq:greenharm}), we can derive a useful identity that connects different time scales:
\begin{equation}
G^0_I(x,0,x',t')=i\int G^0_I(x,0,y,t) G^0_I(y,t,x',t')\mathrm{d}y.
\label{eq:idenity}
\end{equation}
Since a $D$-dimensional harmonic trap decouples in Cartesian coordinates, the corresponding Green's function is just a product of 1D Green's functions from Eq. (\ref{eq:greenharm}).
This product can be rewritten (e.g.~using the Mehler formula) in the following form
\begin{align}
G^0_I(x,0,x',t) = -i e^{i\frac{D\Omega t}{2}}\left(\frac{m_I\Omega}{2i \pi \hbar  \sin(\Omega t)}\right)^{\frac{D}{2}} \nonumber \times \\
\exp\left[\frac{im_I \Omega}{2\hbar \sin(\Omega t)}\left((x^2+{x'}^2)\cos(\Omega t)-2 x x'\right)\right] \; ,
\label{eq:greenharmI}
\end{align}
where the terms of the form $x x'$ should be understood as dot products.
The 1D Green's function for the bosons depleted from the condensate is 
written as 
\begin{equation}
G^0_B(x,0,x',t) = -i\sum_{n=1}^{\infty} \Phi_n(x)
\Phi_n(x')e^{-i\epsilon_nt/\hbar}\; ,
\label{eq:greenharmB}
\end{equation}
where $\Phi_n(x)$ is given by Eq.~(\ref{eq:hermite}) with $m_B$
instead of $m_I$. We see that $G_B^0$ can be written in a form similar to Eq.~(\ref{eq:greenharmI}) 
up to the extra term $i\Phi_0(x)\Phi_0(x')$
since in Eq.~(\ref{eq:greenharmB}) the sum starts from one.
This extra term is constant in a homogeneous case and can be neglected. 
It can also be disregarded if one considers only the leading order perturbation 
from the weak interacting potentials, where one neglects higher order corrections in $\Gamma$. 
Moreover, this term also gives higher order contributions in the full Green's functions if one 
considers short time scales, $\Omega T \ll 1$, since in this case the leading effect arises from 
the propagation in free space. To get rid of this term completely one can redefine $G_B^0$ to also 
include $n=0$ term.
Physically, this corresponds to depleted particles remaining in the ground state of the trap
which propagate with a phase different from the condensate.
In our discussion below, we adopt this reformulation of the Green's function, which now reads
\begin{align}
G^0_B(x,0,x',t) = -i e^{i\frac{D\Omega t}{2}}\left(\frac{m_B\Omega}{2\pi \hbar i \sin(\Omega t)}\right)^{D/2} \times \nonumber \\
\exp\left[\frac{im_B \Omega}{2\hbar \sin(\Omega t)}\left((x^2+{x'}^2)\cos(\Omega t)-2 x x'\right)\right]. 
\end{align}
These expressions can  be easily written 
in the momentum-time domain. For simplicity in the derivations below we will use wave vectors instead of momenta.  We define the corresponding Green's function
as
\begin{equation}
G(p_1,t_1,p_2,t_2)=\int G(x_1,t_1,x_2,t_2)e^{ix_1p_1-ix_2p_2}\mathrm{d}x_1\mathrm{d}x_2.
\end{equation}
The signs in the argument of the exponent are determined from the Fourier transform 
of the creation (annihilation) operators in the definition of the Green's function. 
The Green's function in momentum space for the impurity is 
\begin{align}
G^0_I(k,0,k',t) =-i e^{Di\Omega t/2} \left( \frac{2\pi \hbar}{i m_I \Omega \sin(\Omega t)}\right)^{D/2} \times \nonumber \\
\exp\left[\frac{\hbar }{2 i m_I \Omega \sin(\Omega t)}\left((k^2+{k'}^2)\cos(\Omega t)-2 k k'\right)\right],
\end{align}
and for the bosonic particles the mass $m_B$ should be used instead of $m_I$.

{\it Interaction}.
We would like to investigate universal behavior of the system so any particular form of 
the potential will work. In this paper, we employ a simple Gaussian potential, 
\begin{equation}
V(x)=g\left(\frac{\Lambda}{\sqrt{\pi}}\right)^{D}e^{-x^2 \Lambda^2}, 
\label{eq:gausspot}
\end{equation}
where $g$ and $\Lambda$ are 
parameters that determine the interaction. Note that for $\Lambda \to \infty$ the interaction becomes 
of zero range. For the derivations, we will use this limit as we cannot describe short-distance physics
anyway. In two and three spatial dimensions 
one has to relate the coupling strength $g$ to the scattering length
before taking the zero-range limit
in order to describe physical observables (see Appendix \ref{sec:appa}). 

\section{T-Matrix}
Let us see what happens if the limit $\Lambda\to\infty$ is taken right away which yields $V(x)=g\delta(x)$. We know that in free space 
$\Gamma$ corresponds to a dimer propagator, so we look for $\Gamma$ in the following form 
\begin{widetext}
\begin{align}
\Gamma(x_1,y_1,0,x_2,y_2,t) =
\left(\frac{(m_I+m_B)\Omega}{2\pi \hbar i \sin(\Omega t)}\right)^{\frac{D}{2}} \delta(x_1-y_1) \delta(x_2-y_2)
f(t) e^{Di\Omega t+\left[\frac{i(m_I+m_B) \Omega}{2\hbar \sin(\Omega t)}\left((x_1^2+{x_2}^2)\cos(\Omega t)-2 x_1 x_2 \right)\right]},
\label{eq:selfenergy}
\end{align}
\end{widetext}
where $t>0$. The function $f(t)$ describes 'formation' and 'decay' of this virtual (coherent) dimer. 
The form of $\Gamma$ in Eq.~(\ref{eq:selfenergy}) is obvious by noticing that the frequency is the same for 
both particles. Therefore the product $G_I^0G_B^0$ in Eq.~(\ref{gamma}) can be interpreted as a dimer propagator 
and Eq.~(\ref{eq:idenity}) can be applied. For more complicated trapping potentials the form of $\Gamma$ is less obvious. 

It is worth to note that $f(t)$ determines the time scale at which the diagrams with the T-matrix become important when 
compared to the free propagation. For instance in the example we consider in the next section, 
we use $f(t)=g\delta(t)$, which means that for $t\ll\hbar/(gn)$ the particle can be well 
described using only the free propagator. In the same way by comparing the diagrams that are not in the ladder 
approximation with the diagramms that are included, the function $f(t)$ can be used to validate the use of 
the ladder approximation in a dilute gas. 

The equation for $f$
 can be obtained from Eq. (\ref{gamma}) 
\begin{align}
f(t)=g\delta(t)-\frac{i g}{\hbar} \left(\frac{\mu \Omega}{2\pi \hbar }\right)^{\frac{D}{2}}\int_{\lambda}^{t} 
\left[\frac{-i} {\sin(\Omega\tau)}\right]^{\frac{D}{2}} f(t-\tau)\mathrm{d}\tau,
\label{eq:f}
\end{align}
where $\mu=m_Im_B/(m_I+m_B)$ is the reduced mass and the lower integration limit $\lambda\sim 1/\Lambda^2$ 
denotes that the integral should be regularized as otherwise it appears to be divergent if $D>1$. The need to introduce this regulator is 
related to the use of zero-range interactions for which $g\to 0$ if the scattering length is fixed (see Appendix 
\ref{sec:appa}), 
so the limit of $\Lambda \to \infty$ should be taken properly. This procedure yields finite quantities expressed 
through the scattering length as
shown in Appendix \ref{sec:appb} for 2D and 3D. Equation (\ref{eq:f})
is already suitable for numerical calculations. However, since the equation is linear in $f$ we perform a Fourier transform to obtain further 
analytic insight:
\begin{equation}
f(t)=\frac{1}{2\pi}\int f(\omega)e^{-i \omega t}\mathrm{d}\omega.
\end{equation}
which leads to the equation for $f(\omega)$,
\begin{equation}
f(\omega)=g-\frac{i gf(\omega)}{\hbar} \left(\frac{\mu\Omega}{2\pi \hbar }\right)^{\frac{D}{2}}\int_{\lambda}^{\infty}
\left[\frac{-i} {\sin(\Omega\tau)}\right]^{\frac{D}{2}}e^{i\omega \tau}\mathrm{d}\tau ,
\label{eq:fomegageneral}
\end{equation}
where we used that $f(t)$ is zero for $t<0$. To make sense of the integral in the expression, we imply that $\omega$ has a vanishing positive imaginary part, $\omega\to\omega+i 0$. It is worth noting that one can restrict the integration limits in eq. (\ref{eq:fomegageneral})using periodicity of the $\sin(\Omega t)$ function which might be useful, for example, for regularization of the integral.  Now we examine this equation in different spatial dimensions.

{\it One-dimensional case.}
We first consider the 1D case where we have
\begin{equation}
f(\omega)=g-\frac{i g f(\omega)}{\hbar} \sqrt{\frac{\mu \Omega}{2\pi \hbar}}\int_{0}^{\infty}
\sqrt{\frac{1}{i\sin(\Omega\tau)}}e^{i\omega \tau} \mathrm{d}\tau .
\label{eq:f1d}
\end{equation}
For cold atomic gases experiments in highly elongated "cigar"-shaped optical traps it was shown \cite{olshanii1998} that
\begin{equation}
g=\frac{2\hbar^2 a}{\mu L^2}\frac{1}{1-C\frac{a}{L}},
\end{equation} 
where $a$ is the 3D scattering length,
$L$  is the harmonic oscillator length in the squeezed direction, 
and $C=|\zeta(1/2)|=1.46 ...$, where $\zeta$ is the Riemann zeta
function.
The integral in Eq. (\ref{eq:f1d}) can be easily taken such that
\begin{align}
f(\omega)=g+\frac{  g f(\omega)}{\hbar} \sqrt{\frac{\mu}{4 \pi \hbar\Omega}} B \left(\frac{1}{2},\frac{\omega}{2\Omega}+\frac{1}{4}\right)  \cot\left(\frac{\omega \pi}{\Omega}\right),
\end{align}
where $B(x,y)$ is the standard beta function. For the sake of argument here and below we will assume that $\Omega\to0$, which corresponds to considering short time scales $\Omega t\ll 1$. It is worth to note that the results determined in this way
should have the same form as for the homogeneous case, where $\Omega=0$, so $\Omega t\ll1$ is trivially satisfied. This assumption allows us to demonstrate the main ideas in a relatively simple manner. For practical purposes $f(t)$ can be easily found numerically either by using the inverse Fourier transform or by solving the corresponding equation directly in the time domain.

Assuming short times leads to
\begin{equation}
f^\delta(\omega)=\frac{g}{1+\frac{i g}{\hbar}\sqrt{\frac{\mu}{2\hbar }}\frac{1}{\sqrt{\omega}}}=g-\frac{i g^2}{\hbar}\sqrt{\frac{\mu}{2\hbar}}\frac{1}{\sqrt{\omega}+\frac{i g}{\hbar}\sqrt{\frac{\mu}{2\hbar }}},
\label{eq:fdelta1D}
\end{equation}
where the superscript $\delta$ means that $\Omega t\ll1$.
The inverse transformation can be easily performed for large and small values of $g$. Indeed if $g\to 0$ we obtain $f^\delta(\omega)\simeq g(1-\frac{ig}{\hbar}\sqrt{\frac{\mu}{2\hbar}}\frac{1}{\sqrt{\omega}})$, which corresponds to
 $f^\delta(t)=g\delta(t)-g^2\frac{1+i}{\hbar}\sqrt{\frac{\mu}{4\hbar \pi}}\frac{1}{\sqrt{t}}$. These terms as well as higher order contributions can be derived directly in the time domain (see Appendix \ref{sec:appc}). 
For repulsive interaction, $g>0$,  one can easily perform inverse Fourier transform 
\begin{equation}
f^{\delta}(t)=g\delta(t)+\frac{i g^2}{2 \hbar \pi} \left[-\frac{ \sqrt{2\mu \pi}}{\sqrt{i \hbar t}} + 
   \frac{g \mu \pi}{\hbar^2} 
    \mathrm{erfcx}\left(\frac{g \sqrt{i \mu t}}{
    \hbar \sqrt{2\hbar}}\right)\right],
\label{eq:fdelta1Dt}
\end{equation}
where $\mathrm{erfcx}(x)$ is the scaled complementary error function.
This form corresponds to the limit of the infinite sum derived in  Appendix \ref{sec:appc}. For strong repulsive interaction, i.e. $g\to\infty$, for very short times, i.e. $t\ll\sqrt{\frac{\hbar^3}{g^2\mu}}$ the function $f^{\delta}(t)$ is the same as for the weakly-interacting case discussed above. For larger times we derive 
\begin{equation}
f^\delta(t)\simeq \hbar\frac{i-1}{t^{3/2}}\sqrt{\frac{\hbar}{4\pi\mu}}.
\end{equation}
Note that, this behavior of $f(t)$ leads to non-analytic behavior of $G_I$ for infinite repulsion. We understand that this shows strong correlations taking place in 1D.  These correlations lead to depletion of the condensate which is not properly taken into account in the present approach. 

{\it Three-dimensional case}.
After proper regularization (see Appendix \ref{sec:appb}) we derive in 3D
\begin{equation}
f^\delta(\omega)=\frac{1}{\frac{\mu}{2\pi \hbar^2 a}+i\left(\frac{\mu}{\hbar}\right)^{3/2}\frac{\sqrt{\omega}}{\sqrt{2}\pi \hbar}}.
\label{eq:fdelta3D}
\end{equation}
Notice that this equation with $\omega\to \omega-\frac{p^2}{2\hbar(m_I+m_B)}$ can be compared with the self energy derived in Ref. \cite{schmidt2013} for the homogeneous case, i.e. with $\Omega=0$, which corresponds to the pair propagator in free space. It is also interesting to note that the functional form of Eq. (\ref{eq:fdelta3D}) is similar to Eq. (\ref{eq:fdelta1D}). Apparently the parametric dependence on $g$ in 1D is the same as dependence on $-1/a$ in 3D, which means, for instance, that the weakly interacting 1D case has the same time dependence as the strongly interacting 3D case. It also means that Eq. (\ref{eq:fdelta1Dt}) with some minor changes in constants can be applied to find the inverse Fourier transform for $a<0$ in 3D.   
Let us discuss the case of $a<0$ now in more details.
Apparently, if $a\to 0$, $f^\delta(t)$ to leading order is a delta function. Obviously, this form is due to neglecting higher order diagrams such that the virtual dimer does not propagate in time. At unitarity when $1/a = 0$, the expression for $f^\delta(t)$ can also be found easily. We note that $f^\delta(\omega)$ has a branch cut for negative values of $\omega$, where we take $f^\delta(\omega)\sim 1/\sqrt{\omega+i\delta}$ as discussed above. This leads to vanishing $f^\delta(t)$ for negative values of $t$. This expression can be easily transformed back to the time domain, where the following form should be used
\begin{equation} 
f^\delta(t)=-(1+i)\sqrt{\pi}\left(\frac{\hbar}{\mu}\right)^{3/2}\frac{\hbar}{\sqrt{t}},
\end{equation}
which up to the delta function mimics time dependence of $f^{\delta}(t)$ for weak interaction in 1D. The case of $a>0$ is more involved, since in 3D it implies existence of both a bound molecular state and repulsive polaron branches, see e.g. Ref. \cite{schmidt2013}. We briefly address this problem for small positive $a$ in the time domain in Appendix \ref{sec:appd}.

{\it Two-dimensional case}. Finally, we briefly mention the 2D case. After regularization (see Appendix
\ref{sec:appb}), we obtain the following expression
\begin{equation}
f^\delta(\omega)=\frac{2\hbar^2\pi}{\mu}\frac{1}{i\pi-\ln(a^2_{2D}\mu \omega e^{\gamma}/\hbar)},
\end{equation}
where $a_{2D}$ is the two-dimensional scattering length, and $\gamma = 0.5772...$ is Euler's constant.
Again, we note that this equation with $\omega\to \omega-\frac{p^2}{2\hbar(m_I+m_B)}$ can be used to determine the 
self-energy without the trap.

\section{Dynamics}

After the form of the T-matrix is established one can approach the wave function for the impurity,
\begin{equation}
I(x,t)=\sum_{n=0}^{\infty} \prod_{i=0}^n (N G_I^0 \Gamma)^i G_I^0 I(t=0),
\label{eq:Ilast}
\end{equation}
where operator notation is assumed. This equation is obtained by combining eq. (\ref{eq-wave-funct}) with the Green's function  from eq. (\ref{green}) written as an iterative sum. In this expression 
the integrals over the space variables can easily be taken, since the functions are just Gaussians and the integrals over time variables can be performed numerically. Notice that higher order terms contribute to longer time evolution and are usually suppressed if one is interested in short-time dynamics.

We turn our attention to a one-dimensional weakly-interacting system where the interaction is switched on at time $t=0$ and the impurity starts to move in the trap. This simple system allows us to illustrate time dynamics almost analytically. In this paper, we are mostly focusing on the short-scale time dynamics, however, a quench of the coupling constant in the Hamiltonian can also be used to improve our understanding of non-equilibrium dynamics and relaxation processes in closed quantum systems, see e.g. \cite{polkovnikov2011,caux2014} and references therein.

The motion of the impurity is described with the function $I(y,t)$ in Eq.~(\ref{eq:Ilast}) where for the sake of argument we assume that the system is weakly-interacting and $f(t)\simeq g\delta(t)$. 
To find $I(x,t)$, we can use Eq.~(\ref{eq:Ilast}) directly. Another way 
is to notice that analog to the free Green's function for the weakly-interacting case, we have
\begin{equation}
G_I(x,t,x',t')=i\int \mathrm{d}y G_I(x,t,y,\tau)G_I(y,\tau,x',t'),
\label{eq:gconnect}
\end{equation}
which allows us to construct the infinitesimal operator of time evolution  
\begin{align}
&G^{\delta}_I(x,t,x',t')=G^0_I(x,t,x',t')+ \nonumber \\
&\qquad g N\int \mathrm{d}y \mathrm{d}\tau \Phi^2_0(y) G^0_I(x,t,y,\tau)  G^0_I(y,\tau,x',t'),
\label{eq:greendelta}
\end{align}
and then apply it repeatedly to the initial wave function.
The integration over spatial coordinates can easily be taken, since only Gaussian functions enter.
We are then left only with the integration over the time variables which we perform numerically.
It is worth to note that Eq.~(\ref{eq:gconnect}) cannot be applied to more strongly interacting systems, 
where correlations between different times are more complicated.

We first discuss the homogeneous case, i.e. $\Omega = 0$. In this case the infinitesimal time evolution operator will be written as 
\begin{equation}
G^{\delta}_I(x,t,x',t')=(1-i (t'-t) g n/\hbar )G^0_I(x,t,x',t'),
\end{equation}
where $n$ is the density of BEC. Notice that this operator does not 
conserve the normalization of the wave function in the order $g^2(t'-t)^2$ due to the infinitesimal character of the operator. After repeated use, 
we obtain the following Green's function
\begin{equation}
G_I(x,t,x',t')=e^{-\frac{i g n (t'-t)}{\hbar}}G^0_I(x,t,x',t'),
\end{equation}
which describes the shift of the energy due to interaction. This form of the Green's function implies that if the impurity is in an
eigenstate, it will remain in the same state with some additional phase oscillations. In particular,
if $I(x,0)$ is an eigenstate, then $I(x,t)=e^{-i g n t/\hbar} I(x,0)$.  This approach can be regarded as a mean-field treatment of the problem, i.e. we assume that the bosons form a homogeneous external field, which is not affected by the impurity.  If the condensate density is not constant then these phase oscillations 
will be position-dependent, leading  to a probability current. One would expect this flux initially to be proportional to the gradient 
of the density of the condensate multiplied by $g I^2(x,0)$. At later times additional fluxes due to the external trap will appear. 
Since in the equations only the combination $g \Phi^2_0(x)$ enters, the same behavior emerges also if the condensate is homogeneous 
but the coupling constant is a function of the coordinates.

\begin{figure}
\centerline{\includegraphics[width=20em]{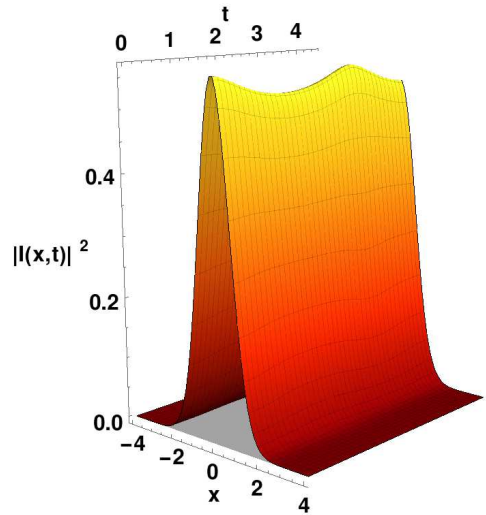}}
\caption{(Color online) Density of the impurity, $|I(x,t)|^2$ as a function of time, $t$, and position coordinate, $x$. 
The interparticle interaction is assumed to be constant in time, $g N =0.3$.  $t$ is in units of $1/ \Omega$, $x$ is in units of $\sqrt{\hbar/(m_I\Omega)}$ and $|I(x,t)|^2$ is in units of $\sqrt{(m_I\Omega)/\hbar}$. }
\label{fig:dens}
\end{figure}

To describe this system, we need to find $G^{\delta}_I(x,t,x',t')$ which now has a more complicated form compared to the homogeneous case. 
However, using simple numerical integration routines the wave function $I(y,t)$ still can be obtained. To simplify the notation, 
we use the units where $\hbar=m_I=\Omega=1$. This corresponds to measuring time in units of $1/\Omega$, energy in units of $\hbar \Omega$, and 
distances in units of $\sqrt{\hbar/(m_I\Omega)}$. For our numerical calculations,  we choose $m_I=m_B$ and $g N=0.3$. 
The corresponding density of the impurity as a function of time is shown in in Fig.~\ref{fig:dens}.

We see that the initial density starts to oscillate with almost complete revival after some time. 
This is easily understood noticing that a weak external potential $gN\Phi^2_0(x)$ acting on the impurity would yield the same diagrams as we use now. 
Apparently, this external potential changes the eigenstates of the Hamiltonian such that the dynamics in the system 
should be described with the following Green's function
\begin{equation}
G_I (x,0,x',t')= -i\sum_{n=0}^\infty F^{*}_n(x)F_n(x')e^{-i\varepsilon_n t/\hbar},
\label{eq:GIP}
\end{equation}
where $\varepsilon_n$ is a new eigenenergy and $F(x)$ is the corresponding eigenfunction.  
Using first order perturbation theory~\cite{landaubook}, we approximate 
these quantities as
\begin{align}
&\varepsilon_n\simeq n+g N \alpha_{nn}, \;
\alpha_{nk} =\int \Psi^{*}_n(x) \Phi^2_0(x) \Psi_k (x)\mathrm{d}x, \nonumber \\
&F_n(x) \simeq \Psi_n(x) + g N \sum_{k\neq n}\frac{\alpha_{kn}}{\epsilon_n-\epsilon_k}\Psi_k(x).
\label{eq:pertfirst}
\end{align}
Assuming $I(x,0)=\Psi_0(x)$, the wave function at later times
up to the phase factor reads
\begin{equation}
I(x,t)=\Psi_0(x)+g N \sum_{n\neq0}\frac{\alpha_{0n}}{n}\Psi_n(x)\left(e^{-i(\varepsilon_n-\varepsilon_0)t/\hbar}-1\right).
\label{eq:Ipert}
\end{equation}
We checked that this wave function reproduces the density from Fig.~\ref{fig:dens} very accurately.
Noticing that $\alpha_{0n}$ in Eq.~(\ref{eq:Ipert}) is non-zero only for even $n$ due to the parity conservation, 
it becomes obvious that after $t=\pi$ the density almost completely returns to its initial profile. This revival is due to the equidistant energy spectrum of the harmonic trap so the period of oscillations is determined by the oscillator frequency. 
The coupling constant and the density of bosons determine the amplitude of the oscillations. 

From previous studies in solid state physics \cite{kenrow1996, ku2007}, it is known that the bare charge carrier evolves into a polaron by emitting phonons. The time dynamics of such a process strongly depends on the parameters of the problem. At the same time Fig.~\ref{fig:dens} indicates that the impurity does not evolve into a polaron at the time scales given by the trap.
This result arises since our model forbids the depletion of the condensate in the final state. In the ideal case, this approximation should be justified. Indeed, there the excitation processes are of one-body nature and thus the depletion processes at the presented time scales are suppressed by $1/\sqrt{N}$ compared to non-depleting processes.  
To arrive at this conclusion, it is crucial  that the interaction between the bosons is absent or very weak. In this context 'very weak' means $N a_{BB} \sqrt{\hbar/(m_I\Omega)}/L^2 \ll  1$ in the quasi-one-dimensional regime. If this condition is satisfied then the system behaves as an ideal gas \cite{stringari2002}.

Even though one can engineer effectively 
non-interacting systems in a laboratory by tuning the boson-boson scattering length~\cite{chin2010}, $a_{BB}$, the question of what happens if the system posesses a weak boson-boson interaction is important. In this case, the boson-boson interaction term, 
\begin{equation}
\frac{1}{2}  \int \mathrm{d}x_1 \mathrm{d}x_2 \phi^\dagger(x_1)\phi^\dagger(x_2) V_{bb}(x_1-x_2) \phi(x_1) \phi(x_2),
\end{equation}
should be added to the Hamiltonian.
If this term is included, the excitation processes are of collective nature, see e.g. Ref. \cite{dalfovo1999, stringari2002}. Then one should take the depletion of the condensate explicitly into account, which will lead to the formation of a polaron by emitting collective excitations. The details of the formation, however, will depend strongly on the strength of the boson-boson interaction and are left for the future studies, i.e., in the present paper we always put $V_{bb}=0$. For relevant studies with $V_{bb}\neq0$ in 1D see Ref.~\cite{Bruderer2012} and an investigation of a three-body ensemble, Ref.~\cite{Campbell2015}. 

\begin{figure}
\centerline{\includegraphics[scale=0.7]{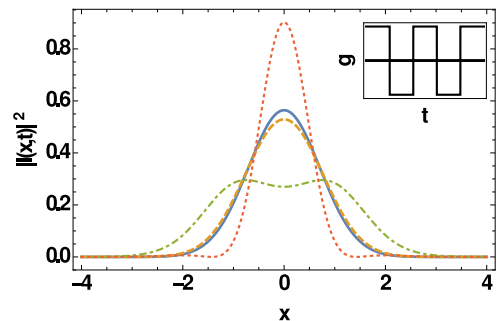}}
\caption{(Color online) Density profiles, $|I(x,t)|^2$, as a function of the coordinate $x$ taken at different times: $t=0$ is shown by the solid (blue) line; 
$t=\pi/2$ - dashed (orange); $t=9\pi/2$ - dot-dashed (green); $t=5\pi$ - dotted (red). The inset shows the interparticle interaction 
which is a periodic function of time with period $\pi$ and peak amplitude $g N=0.3$.  $t$ is in units of $1/ \Omega$, $x$ is in units of $\sqrt{\hbar/(m_I\Omega)}$, $|I(x,t)|^2$ is in units of $\sqrt{(m_I\Omega)/\hbar}$ and $g$ is in units of $\hbar\Omega \sqrt{\hbar/(m_I\Omega)} $}. 
\label{fig:dens1}
\end{figure}

\begin{figure}
\centerline{\includegraphics[width=20em]{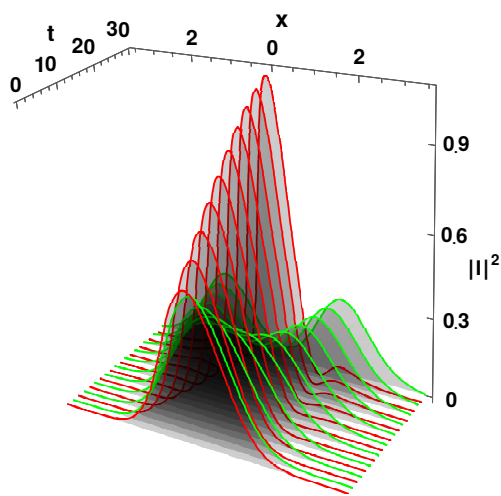}}
\caption{(Color online) Density profiles, $|I(x,t)|^2$, as a function of the coordinate $x$ 
for different times $t=k\pi/2$ with $k\in \{0, 1, 2, ..., 19\}$.  $t$ is in units of $1/ \Omega$, $x$ is in units of $\sqrt{\hbar/(m_I\Omega)}$ and $|I(x,t)|^2$ is in units of $\sqrt{(m_I\Omega)/\hbar}$.  }
\label{fig:densq}
\end{figure}

Figure~\ref{fig:dens} graphically demonstrates that the system has some intrinsic time scale, so if one designs the coupling strength such that it depends on time it can give additional tools to study dynamics. For instance if the interaction is turned off after time $t=\pi$ the oscillations of the density would be reduced, whereas if the interaction is turned off after time $t=\pi/2$ the amplitude of the density oscillations would be enhanced.
To illustrate this we assume that the coupling constant is a periodic function of time as in the inset of Fig.~\ref{fig:dens1}. 
We assume that the period is $\pi$, i.e. the interaction $g N=0.3$ is turned on at $t=0$ then it is constant until $t=\pi/2$ where it changes sign and so forth. Obviously, in the homogeneous case the density of the impurity will not be altered. On the contrary, for the trapped case we should see that the amplitude of the density oscillations increases since the intrinsic time scale given by the harmonic trap coincides with the period for the coupling constant. To investigate this resonant behavior, we use the Green's function from Eq.~(\ref{eq:GIP}) with Eq.~(\ref{eq:pertfirst}), which is easy to implement numerically for long time scales. The result of this investigation is shown in Fig.~\ref{fig:dens1} and Fig.~\ref{fig:densq}. 
In Fig.~\ref{fig:dens1}, we present $|I(x,t)|^2$ at four different times, while Fig.~\ref{fig:densq} shows the evolution of the density from $t=0$
to $t=19\pi/2$.
We clearly see that the amplitude of the oscillations 
at times equal to integer multiples of $\pi$
becomes much larger which means that higher natural orbits of the harmonic 
oscillator are populated.
This demonstrates that even a very weak interparticle interaction appropriately modulated in time can drastically modify 
the initial wave function of the impurity.

\begin{figure}
\centerline{\includegraphics[scale=0.7]{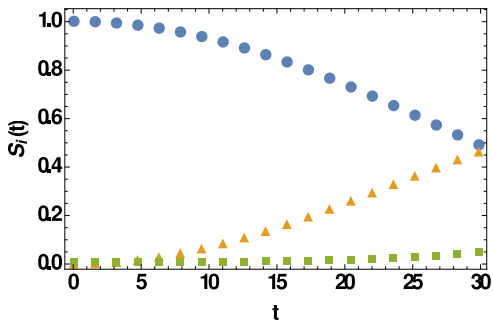}}
\caption{(Color online) The overlap $S_i(t)$ from Eq.~(\ref{eq:Lecho}) as a function of time $t$. 
The (blue) circles correspond to $i=0$, while the (yellow) triangles and (green) squares show the cases $i=2$ and $i=4$, respectively. $t$ is in units of $1/\Omega$. }
\label{fig:overlap}
\end{figure}

To quantify the evolution of the system we calculate the overlap between the states $\Psi_i(x)$ and $I(x,t)$. That is we calculate 
\begin{equation}
S_i(t)=\left|\int \Psi_i(x) I(x,t)\mathrm{d}{x}\right|^2.
\label{eq:Lecho}
\end{equation}
From the definition it is obvious that $S_i(t)$ determines the decomposition coefficients of $I(x,t)$ in the full basis $\Psi_i(x)$. From another point of view $S_0(t)$ compares the evolution of the interacting system with the non-interacting system and it is often referred to as the Loschmidt echo in the literature \cite{perez1984,jalabert2001}.
Since the parity is conserved, $S_{i}(t)=0$ if $i$ is an odd integer.  We present $S_i(t)$ for $i=0,2,4$ in Fig.~\ref{fig:overlap}.
The $S_i$ for $i\geq 6$ are very small for the time scales we consider. The figure clearly shows that the quench protocol from the inset 
in Fig.~\ref{fig:dens1} allows one to reach small values of the Loschmidt echo, $S_0(t)$, 
and enhance $S_{i>0}$ even with weak interactions.

For 2D and 3D a similar investigation for weakly-interacting systems can be provided. However, the dynamics there can be understood from the 1D case. 
Indeed the dynamics in the system should be driven with the Green's function similar to Eq.~(\ref{eq:GIP}) with additional complications arising 
from the degeneracy of the oscillator levels. This complication can be lifted by noticing that angular momentum should be conserved. For example 
if the impurity is initially in the ground state, we expect the corresponding wave function to have zero angular momentum also for $t>0$. 
From the same considerations as above, we conclude that the density oscillations for an interparticle interaction 
constant in time will happen in the radial direction with period $t = \pi$. Since this is very similar to the 1D case, 
we do not go into more details here.

\section{Summary and Outlook}

In this paper, we consider an ideal Bose gas with an impurity in  harmonic trap.
Assuming that the boson-impurity interaction is turned on at time $t=0$, we investigate real-time dynamics of the impurity.
We present a field-theoretical approach to this problem using the ladder approximation. First, 
we show how to calculate the in-medium T-matrix in the space-time domain which form the basis for studying the dynamics of the system. 
As an explicit example, we address a weakly-interacting system in one spatial dimension. We build an infinitesimal time evolution operator 
to investigate the motion of the impurity. By repeated use of this operator, we show that the density of the impurity starts to oscillate 
with a period determined by the trapping frequency. We also show that the amplitude of the oscillations can be enhanced by periodically 
turning the interactions on and off.

A number of interesting questions can be addressed  directly using the ideas presented here. First, one can extend 
our approach to non-isotropic traps in order to address quasi-one (two)-dimensional setups which are often used in
experiments. The corresponding T-matrix can be easily found assuming different trapping frequencies in different directions. 
Second, while our formalism takes into account different masses for the bosons and the impurity, we have concentrated on 
equal masses for our numerical example.
However, different masses would strongly affect the dynamics of the impurity. For instance, 
if the bosons are much lighter than the impurity then the impurity effectively feels a homogeneous distribution of the condensate. For the weakly-interacting case this would lead to absence of density oscillations. On the other hand, if the impurity is very light it should behave like a particle in a harmonic trap with the delta function peak generated by the bosons in the centre \cite{busch1998}.
 Other obvious extensions are the 
effects of stronger interactions and different initial states, i.e. when the impurity initially has a density distribution different 
from the one that was considered.
Naturally, the most interesting extension is to go beyond the diagrams that were included here. 
For example three-body diagrams should lead to some manifestation of Efimov physics~\cite{efimov1970,Braaten:2004rn}. Another interesting extension would be the investigation of the two-point correlation functions in the enviroment. These functions can be used to describe real time correlations between a boson and the impurity or two non-interacting impurities after the boson-impurity interaction quench.

Most of the experiments with Bose gases are carried out with a weak boson-boson interaction.
So a natural extension of the presented work would be to investigate the effect of this interaction on the time dynamics of impurity. Under the assumption that the profile of the condensate is not changed significantly by back action of the impurity one can apply the mean-field treatment to describe the ground state and low-lying excitations of the Bose gas \cite{dalfovo1999}. The real time dynamics of the system then should lead to the depletion of bosons from the 
ground state neglected in this paper.

Our study with an effective single-channel boson-impurity
interaction of zero range
applies to broad Feshbach resonances where a description of the interaction in terms of the scattering length is
sufficient. However, we note that there are also some experiments with narrow Feshbach resonances where the range effects
can be significant \cite{grimm2012}, and therefore should be addressed in the future. 

Finally, we note that even one impurity can significantly modify the density of the Bose gas.
For example, for a very heavy impurity placed in the middle of the trap, one can expect the depletion after time $1/\Omega$ to be very large
if the interaction is sufficiently strong. To describe this effect, one needs to carry out a more sophisticated calculation allowing 
many particles to be depleted from the condensate. Such a calculation would allow one to study the depletion of the condensate in real time.

\begin{acknowledgments}
Discussions with Georg Bruun and Jesper Levinsen are gratefully acknowledged. This work was supported in 
part by Helmholtz Association under contract HA216/EMMI, by the BMBF (grant 06BN9006), and by the Danish Council
for Independent Research DFF Natural Sciences.
\end{acknowledgments}
\appendix

\section{Scattering length in 2D and 3D.}
\label{sec:appa}

{\it Three-dimensional case}.
For short-range potentials  the zero-energy wave function at infinity has the form $\Psi(\mathbf{x})\sim 1-a/|\mathbf{x}|$, which is determined with only one parameter -- the scattering length, $a$.
One can establish \cite{abrikosovbook} the following connection between the scattering length and potential:
\begin{equation}
a=\frac{\mu}{2\pi \hbar^2}\int V(\mathbf{x})\Psi(\mathbf{x})\mathrm{d}\mathbf{x},
\end{equation}
where $\Psi(\mathbf{x})$ is the wave function for the relative coordinate, $\Psi(\mathbf{x\to\infty})=1-\frac{a}{|\mathbf{x}|}$. If the potential is of short range we can approximate this wave function inside of the potential with constant, which allows us to rewrite this definition as
\begin{equation}
a=\frac{\mu}{2\pi \hbar^2}\left(1-\frac{a}{|\mathbf{x_e}|}\right)\int V(\mathbf{x})
\mathrm{d}\mathbf{x},
\end{equation}
where $|\mathbf{x_e}|$ is the range of the interaction. As shown below, for the Gaussian potential considered in the text $|\mathbf{x_e}|=c/\Lambda$ with $c=\sqrt{\pi}/2$ and 
\begin{equation}
a=\frac{g\mu}{2\pi \hbar^2}\left(1-\frac{a \Lambda}{c}\right).
\end{equation}
 Notice that for any non-zero value of scattering length, the limit of $\Lambda\to\infty$ corresponds to $g\to0.$

{\it Two-dimensional case}.
In two spatial dimensions the standard way to define the scattering length, $a_{2D}\geq0$,
for short-range potentials is also through the zero-energy wave function of relative motion, which at infinity should be of the form $\ln(|\mathbf{x}|/a_{2D})$ \cite{verhaar1984}. This form can be obtained directly by solving two-body problem, see for example \cite{volosniev2013}, from which one can deduce
\begin{equation}
\ln(a_{2D}/l)=\frac{-1+\frac{2\mu}{\hbar^2}\int_0^{\infty}\mathrm{d}s s \ln(s/l)V(s)\Psi(s)}{\frac{2\mu}{\hbar^2}\int_0^\infty\mathrm{d}s s V(s)\Psi(s)},
\end{equation}
with $\Psi(0)=1$ and $l$ any parameter with dimension of length. Again assuming that the interaction is of short range, 
we put $\Psi(s)=1$, which for the Gaussian potential leads to the following dependence 
of scattering length on $g$ and cutoff parameter $\Lambda$
\begin{equation}
\ln(a_{2D}\Lambda)=-\frac{\gamma}{2}-\frac{\hbar^2 \pi}{\mu g},
\end{equation}
where $\gamma = 0.57721...$ is the Euler-Mascheroni constant.
Similar to the 3D case $\Lambda\to\infty$ corresponds to $g\to0$.

\begin{widetext}
\section{Regularization procedure for the T-matrix}
\label{sec:appb}
In this Appendix, we show how Eq.~(\ref{eq:f}) is obtained. We also present the corresponding regularization procedure in 2D and 3D. 
Noting that $f(t)$ is independent 
of the spatial coordinates, we show the derivation for the homogeneous case, i.e. $\Omega=0$, which is more transparent.
We first perform a transformation to the momentum-time domain,
since the Green's functions conserve momentum in the homogeneous case, i.e. 
\begin{equation}
G(p,t,p',t') \equiv (2\pi)^D \tilde G(p,t,t')\delta(p-p').
\label{eq:free}
\end{equation}
Moreover, we define the T-matrix in the momentum-time domain as
\begin{align}
\Gamma(p_1,p_1',t_1,p_2,p_2',t_2)=\frac{1}{(2\pi)^D}\int \Gamma(x_1,x_1',t_1,x_2,x_2',t_2)
e^{-ix_1p_1-ix_1'p_1'+ix_2p_2+ix_2'p_2'} \mathrm{d}x_1\mathrm{d}x_1'\mathrm{d}x_2\mathrm{d}x_2',
\end{align}
where $p$ denotes a wave vector.   
With this definition, Eq.~(\ref{gamma}) assumes the following form
\begin{align}
\Gamma(p_1,p_1',t_1,p_2,p_2',t_2)&=V(p_2-p_1)\delta(t_1-t_2)\delta(p_1+p_1'-p_2-p_2')  \nonumber \\
& + \frac{i}{\hbar(2\pi)^{3D}} \int V(p_1'-\kappa)G_I^0(p_1'+p_1-\kappa,t_1,k',t)G_B^0(\kappa,t_1,\kappa',t)\Gamma(k',\kappa',t,p_2,p_2',t_2) \mathrm{d}k'\mathrm{d}\kappa\mathrm{d}\kappa'\mathrm{d}t.
\end{align} 
We rewrite this equation using the Green's function in free space from Eq.~(\ref{eq:free}) 
\begin{align}
\Gamma(p_1,p_1',t_1,p_2,p_2',t_2)=&V(p_2-p_1)\delta(t_1-t_2)\delta(p_1+p_1'-p_2-p_2') \nonumber \\ &+
\frac{i}{\hbar (2\pi)^{D}} \int V(p_1'-\kappa)\tilde G_I^0(p_1'+p_1-\kappa,t_1,t)\tilde G_B^0(\kappa,t_1,t)\Gamma(p_1'+p_1-\kappa,\kappa,t,p_2,p_2',t_2) \mathrm{d}\kappa\mathrm{d}t.
\end{align}
Furthermore, we introduce a new set of variables $P=p_1+p_1', K=p_2+p_2', 2p=p_1-p_1', 2k = p_2-p_2'$. Since the total momentum is conserved, 
we define new function $\tilde \Gamma$ from  $\Gamma(p_1,p_1',t_1,p_2,p_2',t_2) \equiv \delta(P-K)\tilde \Gamma(P;p,k,t_1,t_2)$ which leads to
\begin{align}
\tilde \Gamma(P;p,k,t_1,t_2)=&V(k-p)\delta(t_1-t_2) \nonumber \\ +&
\frac{i}{\hbar (2\pi)^{D}} \int V(P/2-p-\kappa)\tilde G_I^0(P-\kappa,t_1,t)\tilde G_B^0(\kappa,t_1,t)\tilde \Gamma(P; P/2-\kappa, k,t,t_2)\mathrm{d}\kappa\mathrm{d}t.
\end{align}

To illustrate appearance of a pathology, we take $V(k)\equiv g$ where $g$ is a constant satisfying $0<g<\infty$. 
For this potential, the T-matrix reads
\begin{align}
\tilde \Gamma(P;p,k,t_1,t_2)=g\delta(t_1-t_2) + g
\frac{i}{\hbar (2\pi)^{D}} \int \tilde G_I^0(P-\kappa,t_1,t)\tilde G_B^0(\kappa,t_1,t)\tilde \Gamma(P; P/2-\kappa, k,t,t_2)\mathrm{d}\kappa\mathrm{d}t.
\end{align}
Notice that $\tilde \Gamma$ is independent of its second 
and third arguments.
The integral over $\kappa$ can be  taken easily, since the Green's functions are just exponents, so
\begin{align}
\tilde \Gamma(P;t_1,t_2)=g\delta(t_1-t_2) - g
\frac{i}{(2\pi)^{D}\hbar} \int_{t_1}^{t_2} \mathrm{d}t \tilde \Gamma(P;t,t_2)\left(\frac{2\pi \mu }{i \hbar(t-t_1)}\right)^{D/2} e^{-i\frac{\hbar P^2}{2(m_I+m_B)}(t-t_1)}.
\end{align}
The dependence on $P$ can be  eliminated  easily, assuming the following 
form of the T-matrix $\tilde \Gamma(P;t_1,t_2)= e^{-i\frac{\hbar P^2}{2(m_I+m_B)}(t_2-t_1)}f(t_2-t_1)$, 
which leads to the equation for $f$
\begin{align}
f(T)=g\delta(T) - g
\frac{i}{(2\pi)^{D}\hbar} \int_{0}^{T} \mathrm{d}\tau f(T-\tau)\left(\frac{2\pi \mu }{i\tau \hbar}\right)^{D/2} 
\end{align}
This equation is presented and discussed in the main text. However, the equation is pathological. 
For example, the reasonable assumption that $f$ is a non-zero function that is analytic almost everywhere 
cannot be satisfied as the integral has divergence at small times. 

Now we use the Gaussian potential defined in Eq.~(\ref{eq:gausspot})
and take the limit $\Lambda \to \infty$ afterwards. 
Using the results from above, we assume that $\tilde \Gamma$ does not depend on its second and third arguments. 
This yields
 \begin{align}
\tilde \Gamma(P;0,T)=&g\bigg(\delta(T) + 
\frac{i}{(2\pi)^{D}\hbar} \int_{0}^{T} \mathrm{d}t \tilde \Gamma(P;T-t) \int e^{-\frac{\kappa^2}{4\Lambda^2}} \tilde G_I^0(P-\kappa,0,t)G_B^0(\kappa,0,t)\mathrm{d}\kappa \bigg) \nonumber \\ =&
g\bigg(\delta(T) - 
\frac{i}{(2\pi)^{D}\hbar} \int_{0}^{T} \mathrm{d}t \tilde \Gamma(P;T-t)  \left(\frac{\pi}{\frac{1}{4\Lambda^2}+\frac{i t \hbar}{2\mu}}\right)^{D/2} \exp\left[-\frac{\frac{\hbar^2 P^2}{4m_I^2} t^2}{\frac{1}{4\Lambda^2}+\frac{i t\hbar }{2\mu}}-i\frac{\hbar P^2}{2m_I}t \right]\bigg) .
\end{align}
We can neglect value of $1/4$ in the denominator of the argument in the exponent, since it contributes only for very small values of $t$ and exponent is a bounded function. The factor of $1/4$ in the denominator of the fraction cannot be neglected as $(1/t)^{D/2}$ can be an arbitrarily large number for small $t$. 
As before we assume that  $\tilde \Gamma(P;t_1,t_2)= e^{-i\frac{\hbar P^2}{2(m_I+m_B)}(t_2-t_1)}f(t_2-t_1)$ 
which leads to
 \begin{align}
f(T)=
g\bigg(\delta(T) - 
\frac{i}{(2\pi)^{D}\hbar} \int_{0}^{\infty} \mathrm{d}\tau f(T-\tau) \left(\frac{\pi}{\frac{1}{4\Lambda^2}+\frac{i\tau\hbar}{2\mu}}\right)^{D/2} \bigg),
\end{align}
where we extended the integration limits to infinity since the function $f$ should vanish for negative values of the argument.
Notice that assumption that $f$ is analytic almost everywhere and non-zero can be satisfied now since the divergence of integral for $\Lambda\to\infty$ is compensated by the coupling constant $g$, which goes to zero.
It is simpler to take the limit of zero-range interaction after the expression is Fourier transformed to the frequency domain, where we obtain
 \begin{align}
f(\omega)=
g\bigg(1 - 
\frac{i}{(2\pi)^{D}\hbar} f(\omega) \int_{0}^{\infty} \mathrm{d}\tau e^{i\omega \tau} \left(\frac{\pi}{\frac{1}{4\Lambda^2}+\frac{i\tau \hbar}{2\mu }}\right)^{D/2} \bigg).
\end{align}
The integral can be carried out and has the same kind of divergence as $1/g$, 
so the limit $\Lambda\to\infty$ yields a finite expression that depends only on the scattering length. For instance 
in 3D, we have 
\begin{align} 
(1-a\Lambda/c)f(\omega)=\frac{2\pi\hbar^2}{\mu}a\bigg(1 - 
\frac{i}{(2\pi)^{3}\hbar^2} \pi^{3/2}f(\omega) (-8 i \mu \Lambda + 4\sqrt{2\pi\omega}\mu^{3/2} ) \bigg).
\end{align}
This equation can be used to determine $c$ which eliminates the divergence proportional to $\Lambda$ so $c=\sqrt{\pi}/2$.  After this divergence is eliminated we are left with the equation for $f(\omega)$, presented in the main text.
In 2D, we have instead
 \begin{align}
-\frac{\mu}{\hbar^2 \pi}(\ln(a_{2D}\Lambda)+\gamma/2)f(\omega)=
1 + 
\frac{\mu}{2\pi\hbar^2} f(\omega)  \left(\ln \left(\frac{\mu \omega}{2 \Lambda^2 \hbar}\right)+\gamma -i \pi \right),
\end{align}
which yields the result shown above. 

\section{$\tilde \Gamma(0;0,t)$ in the one-dimensional case}
\label{sec:appc}

To obtain $f(T)$ in the main text, we applied the inverse Fourier transform.
In this appendix, we show another way to derive $f(T)$. 
We focus on  1D, since there regularization of the integrals is not needed, by considering $f(t)=\tilde \Gamma(0;0,t)$, 
which is written as 
\begin{align}
\tilde \Gamma(0;0,T)=&g\bigg(\delta(T) + 
\frac{ig}{2\pi\hbar} \int_{0}^{T} \mathrm{d}t \tilde \Gamma(0;T-t) \tilde G_I^0(\kappa,0,t)G_B^0(\kappa,0,t)\mathrm{d}\kappa \bigg).
\end{align}
By definition this integral is understood as the following sum
\begin{align}
\tilde \Gamma(0;0,T)=&g\bigg(\delta(T) + 
\sum_{n=1}^{\infty}\left(\frac{ig}{2\pi\hbar}\right)^n \int_{0}^{T} \mathrm{d}t_1\int_{t_1}^T\mathrm{d}t_2 ...\int_{t_{n-2}}^T\mathrm{d}t_{n-1} \prod_{j=1}^n \int \mathrm{d}\kappa_j \tilde G_I^0(\kappa_j,t_{j-1},t_j)G_B^0(\kappa_j,t_{j-1},t_j) \bigg),
\end{align}
where $t_0=0,t_n=T$; for $n=1$ it is implied that the integral over time is absent. The integrals over $\kappa_j$ are easily taken, such that
\begin{align}
\tilde \Gamma(0;0,T)=&g\bigg(\delta(T) + 
\sum_{n=1}^{\infty}\left(\frac{-ig}{2\pi\hbar}\right)^n \left({\frac{2\mu \pi}{i \hbar}}\right)^{n/2} \int_{0}^{T} \mathrm{d}t_1\int_{t_1}^T\mathrm{d}t_2 ...\int_{t_{n-2}}^T\mathrm{d}t_{n-1} \prod_{j=1}^n \sqrt{\frac{1}{(t_j-t_{j-1})}}\bigg).
\label{eq:G00T}
\end{align}
From dimensional analysis the multidimensional integral should be equal to $a_n T^{n/2-1}$. To find the coefficients, $a_n$, we establish the recursive relation
\begin{equation}
a_nT^{n/2-1}=\int_{0}^{T} \mathrm{d}t_1\int_{t_1}^T\mathrm{d}t_2 ...\int_{t_{n-2}}^T\mathrm{d}t_{n-1} \prod_{j=1}^n \sqrt{\frac{1}{(t_j-t_{j-1})}} = 
\int_{0}^{T} \frac{\mathrm{d}t_1}{\sqrt{t_1}}a_{n-1}(T-t_1)^{n/2-3/2}=a_{n-1}T^{(n-1)/2} B(\frac{1}{2},\frac{n-1}{2}),
\end{equation}
and obtain $a_n=a_{n-1}B(\frac{1}{2},\frac{n-1}{2})$. Since $a_1=1$ we obtain $a_{2k}=\frac{\pi^k}{(k-1)!}$, and $a_{2k+1}=\frac{\pi^k}{(k-1)!}B(\frac{1}{2},k)$. For large values of $k$ we have $B(1/2,k)\simeq \sqrt{\pi/k}$. This allows us to conclude that the coefficients for large $k$ decay roughly as $\pi^{n/2}/(n/2-1)!$, which implies that the sum in Eq. (\ref{eq:G00T}) is convergent for any 
$g$ and $T>0$.  
Finally this sum is written as
\begin{align}
\tilde \Gamma(0;0,T)=&g\bigg(\delta(T) + 
\frac{1}{T}\sum_{n=1}^{\infty}a_n\left(\frac{-ig}{2\pi\hbar}\right)^n \left({\frac{2\mu \pi T}{i \hbar}}\right)^{n/2} \bigg).
\end{align}
We notice that this expression for $g>0$ converges to Eq. (\ref{eq:fdelta1Dt}). 
For 2D and 3D similar ideas can be used, however, a proper regularization procedure is needed 
which makes the calculations more complicated. 

\section{Molecular branch in 3D.}
\label{sec:appd}

In 3D $a>0$ implies existence of a bound molecule, which then smootly evolves to a Bose polaron \cite{schmidt2013}. This smooth evolution should be contrasted with the Fermi polaron case \cite{prokofev2008}.  In the frequency domain the molecule is seen through the pole close to the integration contour, which leads to existence of a double solution of the standard equation for the energy of a quasiparticle
\begin{equation}
\omega=n \mathrm{Re} f(\omega).
\label{eq:quasi}
\end{equation}
It is interesting to note that the double solutions also exist in 1D for $g<0$,
where $4\omega=-\left(g^2\mu \pm g^{3/2}\sqrt{\mu}\sqrt{g\mu-8 n}-4gn\right)$. Note that in this section we put $\hbar=1$ for simplicity. 
We also note that, in 1D for $g>0$ Eq. (\ref{eq:quasi}) not always can be used, at least within the T-matrix approximation, to define a quasiparticle due to strong correlations happening in 1D. 
 In this section we would like to investigate appearence of this molecular solution in 3D for small postitive $a$ in the time domain. For this we write the equation for $f$ for $\mu=1$ using the derivations from Appendix \ref{sec:appb}
\begin{equation}
f(t)=g\delta(t)-ig\left(\frac{1}{2\pi}\right)^{3/2}\int_{0}^t \left(\frac{1}{i\tau+\frac{1}{2\Lambda^2}}\right)^{3/2}f(t-\tau)\mathrm{d}\tau.
\end{equation}
 For very large times this equation reads
\begin{equation}
f(t)=-ig\left(\frac{1}{2\pi i}\right)^{3/2}\int_{0}^{t\to\infty} \left(\frac{1}{i\tau+\frac{1}{2\Lambda^2}}\right)^{3/2}f(t-\tau)\mathrm{d}\tau,
\end{equation}
with one obvious solution $f(t\to\infty)=e^{\frac{i t}{2 a^2}}$ that describes propagation of a molecule. So we look for $f$ in the form $f(t)=K(t)e^{\frac{i t}{2 a^2}}$, where $K(t)$ solves the following equation
\begin{equation}
K(t)=g\delta(t)-ig\left(\frac{1}{2\pi}\right)^{3/2}\int_{0}^t \left(\frac{1}{i\tau+\frac{1}{2\Lambda^2}}\right)^{3/2}K(t-\tau) e^{-i\tau/(2a^2)}\mathrm{d}\tau.
\end{equation}
Let us assume that $K$ is constant. With this assumption we obtain
\begin{align}
K=2\pi a \delta(T)-\frac{i a K}{\sqrt{2\pi}}\left[2\sqrt{\frac{i}{T}}e^{-\frac{iT}{2a^2}}+2i\sqrt{\frac{\pi}{2a^2}}\mathrm{erf}\left(\sqrt{\frac{i T}{2a^2}}\right)\right],
\end{align}
where $\mathrm{erf}(x)$ is the error function. The equation is satisfied for small $a$ if $K=-i\pi/a$. This means that in the limit of small scattering length propagation of the virtual dimer is determined by the following function  $f(t)=-\frac{i \pi}{a} e^{it/(2a^2)}\simeq 2\pi a\delta(t)$. This result is very obvious, since it also should be reached if the interparticle interaction is completely repulsive, such that a molecule cannot be formed.

\end{widetext}

\bibliographystyle{apsrev4-1}
\bibliography{bib}

 \end{document}